\documentclass[twocolumn,showpacs]{revtex4}
 \usepackage{amssymb} \usepackage{graphicx}

\begin{document}
 \title{Couplings of gravitational currents with Chern-Simons gravities}
 \author{\"Umit Ertem}
 \email{umitertemm@gmail.com}
\author{\"{O}zg\"{u}r A\c{c}{\i}k}
\email{ozacik@science.ankara.edu.tr}
\address{Department of Physics,
Ankara University, Faculty of Sciences, 06100, Tando\u gan-Ankara,
Turkey\\}

\date{\today}

\begin{abstract}
The coupling of conserved $p$-brane currents with non-Abelian gauge
theories is done consistently by using Chern-Simons forms. Conserved
currents localized on $p$-branes that have a gravitational origin
can be constructed from Killing-Yano forms of the underlying
spacetime. We propose a generalization of the coupling procedure
with Chern-Simons gravities to the case of gravitational conserved
currents. In odd dimensions, the field equations of coupled
Chern-Simons gravities that describe the local curvature on
$p$-branes are obtained. In special cases of three and five
dimensions, the field equations are investigated in detail.
\end{abstract}


\maketitle

\section{Introduction}

Couplings with external sources in gauge theories are described by
the well-known minimal coupling procedure. However, this is relevant
only for the external point sources, and in the case of extended
objects the situation is different. Extended objects that have
\emph{p} space dimensions are called \emph{p}-branes and they have
(\emph{p}+1)-dimensional worldvolumes. \emph{p}-branes are the
generalizations of the point particles to higher dimensions. Charges
that are localized on \emph{p}-branes define conserved currents in
spacetime. The coupling of these currents with non-Abelian gauge
theories in the standard minimal coupling procedure is problematic
\cite{Teitelboim, Henneaux Teitelboim}. However, in the case of
Chern-Simons (CS) gauge theories, the problem of coupling with
extended sources has a natural solution.

CS theories of non-Abelian gauge fields are metric-free and
background-independent gauge theories that exist in odd dimensions.
CS theories of gravity are also defined in odd dimensions by using
the de Sitter (or anti-de Sitter) gauge connections in the
first-order formalism of gravity \cite{Deser Jackiw Templeton,
Achucarro Townsend, Zanelli1}. The coupling of extended sources with
CS gauge theories generalizes the minimal coupling procedure by
using the transformation properties of CS forms under gauge
transformations. CS forms transform as Abelian gauge connections and
this property produces a consistent gauge-invariant coupling between
a conserved current and a CS form. The interaction term in the
action is gauge invariant up to a boundary term if the coupling
current is conserved. So the conserved currents on 2\emph{p}-branes
can couple with CS gravities consistently. However, CS theories can
only be defined in odd dimensions and hence only the coupling of
even-dimensional branes can be written. Even-dimensional branes and
odd-dimensional CS forms define a consistent interaction term in the
action. One example of conserved currents are the electromagnetic
currents that are defined on \emph{p}-brane worldvolumes. The
coupling between electromagnetic extended sources and CS gravities
were recently studied in the literature \cite{Miskovic Zanelli,
Edelstein Garbarz Miskovic Zanelli}. The coupling of extended
charged events with CS theories was also considered in Ref.
\cite{Bunster Gomberoff Henneaux}.

Conserved charges in theories of gravitation can be defined from the
asymptotical symmetries of the spacetime. Killing vector fields of
asymptotically flat or asymptotically AdS spacetimes are used in
defining the mass and angular momentum in general relativity
\cite{Arnowitt Deser Misner}. On the other hand, for extended
objects like \emph{p}-branes, the definition of conserved currents
can be generalized by using Killing-Yano (KY) forms \cite{Kastor
Traschen, Acik Ertem Onder Vercin1}. KY forms define hidden
symmetries of spacetime that are generalizations of Killing vector
fields to higher-order forms \cite{Acik Ertem Onder Vercin2}.
Conserved currents that are constructed from KY forms are localized
on \emph{p}-branes and conserved charges for these branes can be
defined by using the asymptotical symmetries of transverse
directions to the brane. The conservation of the currents
constructed from KY forms are shown in Ref. \cite{Acik Ertem Onder
Vercin1}.

Conserved gravitational currents can also consistently couple with
CS gravities. Currents localized on \emph{p}-branes affect the local
geometry of the brane and the field equations of CS gravities
coupled with gravitational currents give this local geometry. CS
gravities that have a global AdS structure induce conserved currents
on \emph{p}-branes from KY forms of the AdS spacetime. Because CS
gravities are defined in odd dimensions, the KY forms that have odd
form degrees are used in the construction of these currents. In this
work, we generalize the coupling of conserved currents with CS
gravities to the gravitational-currents case. We find the field
equations that define the local geometry of 2\emph{p}-branes and
give special examples in three and five dimensions.

The paper is organized as follows. In Sec. II we review the
electromagnetic current couplings of 2\emph{p}-branes with CS
gravities. The definition of KY forms and the construction of
conserved gravitational currents from KY forms are included in Sec.
III. The couplings of gravitational currents with CS gravities in
arbitrary odd dimensions are considered in Sec. IV. Section V
presents special examples for three and five dimensions and the
conclusion is given in Sec. VI.

\section{Brane Couplings in CS Theories}

The coupling of conserved currents with gauge connections in
\emph{n} dimensions is provided by the minimal coupling term in the
action
\begin{eqnarray}
I_{MC}=\int_{M^n}d^nxj_{\mu}A^{\mu}
\end{eqnarray}
where $j_{\mu}$ is the conserved current generated by a point source
and $A^{\mu}$ is the vector potential. The generalization of the
minimal coupling procedure to extended sources like \emph{p}-branes
is possible for Abelian connections in the form
$j_{\mu_1\mu_2...\mu_p}A^{\mu_1\mu_2...\mu_p}$. However, for
non-Abelian connections this procedure is not well defined
\cite{Teitelboim, Henneaux Teitelboim}. In the general case, the
couplings of extended objects are described gauge-invariantly by
using CS forms. A $p$-brane is defined as an object that extends to
$p$ space dimensions and has a $(p+1)$ dimensional worldvolume. The
currents localized on $p$-brane worldvolumes are defined by the
transverse directions to the brane. Hence, in $2n+1$ dimensions, the
current localized on a 2$p$-brane is a $2n+1-(2p+1)=2(n-p)$-form.

Let $\textbf{A}$ be a non-Abelian gauge connection that is a Lie
algebra-valued 1-form. The connection transforms under gauge
transformations as follows:
\begin{eqnarray}
\textbf{A}\rightarrow\textbf{A}'=g^{-1}\textbf{A}g+g^{-1}dg
\end{eqnarray}
where $g$ is an arbitrary element of the Lie group. In $2n+1$
dimensions, CS forms are defined from the connection $\textbf{A}$ as
\cite{Nakahara}
\begin{eqnarray}
\langle {\cal{C}}_{2n+1}(\textbf{A})\rangle&=&\frac{1}{n+1}\langle
\textbf{A}(d\textbf{A})^n\nonumber\\
&&+c_1\textbf{A}^3(d\textbf{\textbf{A}})^{n-1}+...+c_n\textbf{A}^{2n+1}\rangle
\end{eqnarray}
where $\langle\quad\rangle$ denotes the invariant symmetric trace,
namely the Cartan-Killing form in the Lie algebra that takes traces
of the Lie algebra elements in the adjoint representation and
$\textbf{A}^n=\textbf{A}\wedge...\wedge\textbf{A}$ (\emph{n} times).
$c_1,...,c_n$ are dimensionless coefficients determined by the
condition
\begin{eqnarray}
d\langle{\cal{C}}_{2n+1}(\textbf{A})\rangle=\frac{1}{n+1}\langle
\textbf{F}^{n+1}\rangle.
\end{eqnarray}
Here $d$ is the exterior derivative operator and
$\textbf{F}=d\textbf{A}+\textbf{A}\wedge\textbf{A}$ is the curvature
of the connection $\textbf{A}$. CS forms transform under gauge
transformations as Abelian connections \cite{Zanelli1},
\begin{eqnarray}
{\cal{C}}_{2p+1}(\textbf{A}')\rightarrow{\cal{C}}_{2p+1}(\textbf{A})+d\Omega_{(2p)}
\end{eqnarray}
where $p=0,...,n$ and $\Omega_{(2p)}$ is an arbitrary 2$p$-form.
This property is responsible for the consistent coupling between
conserved currents and CS forms, because the coupling term in the
action is
\begin{eqnarray}
I_C=\int\langle j_{(2n-2p)}\wedge{\cal{C}}_{2p+1}(\textbf{A})\rangle
\end{eqnarray}
and remains gauge invariant up to a boundary term. Here $j_{(2n-2p)}$ is
a conserved current localized on a 2$p$-brane.

The action of CS gauge theories in $2n+1$ dimensions is defined as
follows:
\begin{eqnarray}
I_{CS}=\kappa\int_{M^{2n+1}}\langle{\cal{C}}_{2n+1}(\textbf{A})\rangle
\end{eqnarray}
where $\kappa$ is a dimensionless constant. A CS theory can couple
with a conserved ($2n-2p$)-form current localized on a 2$p$-brane.
The total action for CS theory coupled with a 2$p$-brane is
\begin{eqnarray}
I_{2n+1}=\kappa\int_{M^{2n+1}}\langle{\cal{C}}_{2n+1}(\textbf{A})-j_{(2n-2p)}\wedge{\cal{C}}_{2p+1}(\textbf{A})\rangle.
\end{eqnarray}
The field equations
\begin{eqnarray}
\textbf{F}^n=j_{(2n-2p)}\wedge\textbf{F}^p
\end{eqnarray}
can be found by varying the action with respect
to $\textbf{A}$. Thus, outside the worldvolume of the brane the field equations are
$\textbf{F}^n=0$. However, on the worldvolume, different solutions
appear. For example, an electromagnetic current on a 2$p$-brane can
be defined as
\begin{eqnarray}
j_{(2n-2p)}=q_{2p}\delta(T^{2n-2p})d\Omega^{2n-2p}\textbf{G}^{\textbf{J}_1...\textbf{J}_{n-p}}
\end{eqnarray}
where $q_{2p}$ is the electric charge on the brane,
$\delta(T^{2n-2p})$ denotes the localization of the current on the
transverse directions $T^{2n-2p}$ to the brane, and
$d\Omega^{2n-2p}$ is the volume form on the transverse directions to
the brane. $\textbf{G}^{\textbf{J}_1...\textbf{J}_{n-p}}$ is
constructed from the Lie algebra generators
$\textbf{J}_1,...,\textbf{J}_{n-p}$. Hence, the current is written
as a Lie algebra-valued $2(n-p)$-form \cite{Miskovic Zanelli}. This
conserved current defines a nontrivial curvature on the brane
through the field equations.

\section{KY Forms and Gravitational Currents}

Conserved quantities in gravitational theories are described by the
symmetries of the underlying spacetime. If the spacetime has Killing
vector fields, which generate local isometries of the spacetime,
then a conserved current can be constructed by using them. The
well-known gravitational 1-form current is written as
$j_{(1)}=K_a*^{-1}G^a$, where $K_a$ are the components of a Killing
vector field $K$, $*^{-1}$ is the inverse Hodge map on differential
forms and $G^a$ are the Einstein $(n-1)$-forms in $n$ dimensions.
Corresponding conserved charges are defined from the asymptotical
symmetries of the spacetime. The generalization of gravitational
conserved currents can be obtained by using KY forms, which
generalize the Killing vector fields to higher-order forms.

If $\omega_{(p)}$ is a KY $p$-form then it satisfies the equation
\begin{eqnarray}
\nabla_{X}\omega_{(p)}=\frac{1}{p+1}i_{X}d\omega_{(p)}
\end{eqnarray}
for all vector fields $X$, which is the generalization of Killing's
equation. Here $\nabla_X$ is the covariant derivative and $i_X$ is
the interior derivative (contraction) operator with respect to the
vector field $X$. This equation implies that all KY forms are
co-closed, namely $\delta\omega_{(p)}=0$, where
$\delta=(-1)^{p}*^{-1}d*$ is the co-derivative operator and * is the
Hodge map on differential forms. For a class of spherically
symmetric spacetimes, solutions of the KY equation in four
dimensions are found in Ref. \cite{Acik Ertem Onder Vercin2}.

Two basic conserved gravitational currents can be defined from the
curvature characteristics and KY forms $\omega_{(p)}$ of the
underlying spacetime. The first current is defined as
\begin{eqnarray}
{\cal{J}}_1(\omega_{(p)})&=&i_{X_a}(i_{X_b}\omega_{(p)}\wedge R^{ba})\\
&=&-i_{X_a}i_{X_b}\omega_{(p)}\wedge R^{ab}+(-1)^p
i_{X_a}\omega_{(p)}\wedge P^a\nonumber
\end{eqnarray}
and the second one is
\begin{eqnarray}
{\cal{J}}_2(\omega_{(p)})&=&(-1)^p i_{X_a}(\omega_{(p)}\wedge P^a)\nonumber\\
&=&(-1)^p i_{X_a}\omega_{(p)}\wedge P^a+{\cal{R}}\omega_{(p)}
\end{eqnarray}
where $R^{ab}$ are curvature 2-forms, $P^a=i_{X_b}R^{ba}$ are Ricci
1-forms and ${\cal{R}}=i_{X_a}P^a$ is the curvature scalar with
${X_a}$ being an arbitrary frame basis. We will use ${\cal{J}}_1$
and ${\cal{J}}_2$ instead of ${\cal{J}}_1(\omega_{(p)})$ and
${\cal{J}}_2(\omega_{(p)})$ for brevity. As was shown in Ref.
\cite{Acik Ertem Onder Vercin1}, both of these currents are
co-closed,
\begin{eqnarray}
\delta{\cal{J}}_1=0=\delta{\cal{J}}_2
\end{eqnarray}
and hence the currents $*{\cal{J}}_1$ and $*{\cal{J}}_2$ are
conserved.

The term "gravitational currents" indeed means that they are defined
from curvature characteristics and hidden symmetries of the
background spacetime, and there is no direct relation between the
currents and the Einstein field equations. Hence, they can be seen
as analogous to the electromagnetic currents in some sense, though
they are different by their way of construction. So, these currents
can be interpreted as they are localized on $p$-branes and can
define charge densities for $p$-brane spacetimes. This opens the
possibility of the coupling of gravitational conserved currents on
$p$-branes with CS gravities.

As a special case, the currents have more simple forms in
constant-curvature spacetimes. The curvature characteristics of an
$n$-dimensional constant-curvature spacetime are given by the
following equalities:
\begin{eqnarray}
R^{ab}&=&ce^a\wedge e^b\nonumber\\
P^a&=&c(n-1)e^a\\
{\cal{R}}&=&cn(n-1)\nonumber
\end{eqnarray}
where $c$ is a constant. Hence the currents defined in Eqs. (12) and
(13) can be written as constant multiples of KY $p$-forms,
\begin{eqnarray}
{\cal{J}}_1&=&-cp(n-p)\omega_{(p)}\\
{\cal{J}}_2&=&c(n-1)(n-p)\omega_{(p)}.
\end{eqnarray}
Thus they are linearly dependent and the conservation of their Hodge
duals is a result of the co-closedness of KY forms. In an
$n$-dimensional spacetime, the maximal number of KY $p$-forms is
given by the number
\begin{eqnarray}
C\left(n+1,p+1\right)=\frac{(n+1)!}{(p+1)!(n-p)!}
\end{eqnarray}
and this number is attained in constant-curvature spacetimes. Hence,
the number of KY $p$-forms in constant-curvature spacetimes gives
the number of independent gravitational conserved currents
constructed from KY $p$-forms.

\section{Couplings of KY Currents with CS Gravities}

In the first-order formalism of gravity, the fundamental fields that
describe gravitational interactions are the co-frame 1-forms $e^a$
and the connection 1-forms $\omega^{ab}$. In $n+1$ dimensions these
two quantities can be combined into a single Lie algebra-valued
gauge connection to construct the AdS (SO$(n-1,2)$) (or dS
(SO$(n,1)$)) gauge theories of gravity \cite{Wise},
\begin{eqnarray}
\textbf{A}=\frac{1}{2}\omega^{ab}\textbf{J}_{ab}+\frac{1}{l}e^a\textbf{J}_a
\end{eqnarray}
where $a,b=0,1,...,n$ and $l$ is a constant in units of length.
$\textbf{J}_{ab}$ and $\textbf{J}_a=\textbf{J}_{an}$ are the
generators of the AdS algebra. The associated gauge curvature 2-form
is written in terms of Riemann curvature 2-forms
$R^{ab}=d\omega^{ab}+{\omega^a}_c\wedge\omega^{cb}$ and torsion
2-forms $T^a=de^a+{\omega^a}_b\wedge e^b$ as,
\begin{eqnarray}
\textbf{F}&=&d\textbf{A}+\textbf{A}\wedge\textbf{A}\nonumber\\
&=&\frac{1}{2}\left(R^{ab}+\frac{1}{l^2}e^a\wedge
e^b\right)\textbf{J}_{ab}+\frac{1}{l}T^a\textbf{J}_a.
\end{eqnarray}
The flat connection $\textbf{F}=0$ corresponds to torsion-free,
constant-curvature AdS spacetime: $R^{ab}=-\frac{1}{l^2}e^a\wedge
e^b$. From now on we take the torsion to be zero.

CS gravities are defined from the AdS connection in $2n+1$
dimensions, and the action that includes a coupling term with a
current localized on a 2$p$-brane is written as in Eq. (8),
\begin{eqnarray}
I_{2n+1}=\kappa\int_{M^{2n+1}}\langle{\cal{C}}_{2n+1}(\textbf{A})-j_{(2n-2p)}\wedge{\cal{C}}_{2p+1}(\textbf{A})\rangle\nonumber.
\end{eqnarray}
From the field equations (9) of this action,
\begin{eqnarray}
\textbf{F}^n=j_{(2n-2p)}\wedge\textbf{F}^p\nonumber,
\end{eqnarray}
it can be seen that in spacetime regions out of the brane, the field
equations have the form $\textbf{F}^n=0$ and the solutions give the
global structure of the spacetime (one solution is $\textbf{F}=0$
which implies the global AdS structure and this equation can also be
satisfied by decomposable \textbf{F}'s). This global structure
defines the conserved gravitational currents localized on
2$p$-branes. This property resembles Mach's principle in gravitation
theories \cite{Barbour Pfister}, which states that the local motion
of a body is determined by the large-scale distribution of matter.
So, the currents defined in Eqs. (12) and (13) are constructed from
curvature characteristics and KY forms of source-free regions of
spacetime.

For the current ${\cal{J}}_1$ defined in Eq. (12), the action
becomes
\begin{eqnarray}
I_{2n+1}&=&\kappa\int_{M^{2n+1}}\langle{\cal{C}}_{2n+1}(\textbf{A})-*{\cal{J}}_1\wedge{\cal{C}}_{2p+1}(\textbf{A})\rangle\nonumber\\
&=&\kappa\int_{M^{2n+1}}\langle{\cal{C}}_{2n+1}(\textbf{A})\nonumber\\
&-&*(i_{X_a}(i_{X_b}\omega_{(2p+1)}\wedge
R^{ba}_{G}))\wedge{\cal{C}}_{2p+1}(\textbf{A})\rangle
\end{eqnarray}
and the field equations are
\begin{eqnarray}
\textbf{F}^n=*(i_{X_a}(i_{X_b}\omega_{(2p+1)}\wedge
R^{ba}_{G}))\wedge\textbf{F}^p
\end{eqnarray}
where the KY forms $\omega_{(2p+1)}$ and curvature 2-forms
$R^{ab}_{G}$ in the current are the characteristics of the global
spacetime. From the definition of the gauge curvature 2-form in Eq.
(20), the wedge product of two curvature forms is
\begin{eqnarray}
\textbf{F}\wedge\textbf{F}&=&\frac{1}{4}\left(R^{ab}+\frac{1}{l^2}e^a\wedge
e^b\right)\nonumber\\
&&\wedge\left(R^{cd}+\frac{1}{l^2}e^c\wedge
e^d\right)\left[\textbf{J}_{ab},\textbf{J}_{cd}\right]\nonumber
\end{eqnarray}
and the field equations are written as follows:
\begin{eqnarray}
&&\frac{1}{2^n}\left(R^{ab}+\frac{1}{l^2}e^a\wedge
e^b\right)\nonumber\\
&&\underbrace{\wedge ...
\wedge}_{n-1}\left(R^{kl}+\frac{1}{l^2}e^k\wedge
e^l\right)\left[\textbf{J}_{ab},...,\textbf{J}_{kl}\right]\nonumber\\
&&=*(i_{X_a}(i_{X_b}\omega_{(2p+1)}\wedge
R^{ba}_{G}))\textbf{J}_{12}...\textbf{J}_{(n-p-1)(n-p)}\nonumber\\
&&\wedge\frac{1}{2^p}\left(R^{ab}+\frac{1}{l^2}e^a\wedge
e^b\right)\nonumber\\
&&\underbrace{\wedge ...
\wedge}_{p-1}\left(R^{pq}+\frac{1}{l^2}e^p\wedge
e^q\right)\left[\textbf{J}_{ab},...,\textbf{J}_{pq}\right]\nonumber
\end{eqnarray}
where $\left[\textbf{J}_{ab},...,\textbf{J}_{kl}\right]$ denotes the
commutator of Lie algebra generators, which comes from the wedge
product of Lie algebra-valued forms.

For the current ${\cal{J}}_2$ defined in Eq. (13), the field
equations become
\begin{eqnarray}
\textbf{F}^n=*(-i_{X_a}(\omega_{(2p+1)}\wedge
P^a_{G}))\wedge\textbf{F}^p
\end{eqnarray}
and the same procedure applies for the multiple wedge products of
gauge curvature 2-forms, as in the first case.

Linear combinations of two currents are also conserved and they can
couple with CS gravity. The field equations are found from the
following action:
\begin{eqnarray}
I_{2n+1}&=&\kappa\int_{M^{2n+1}}\langle{\cal{C}}_{2n+1}(\textbf{A})\nonumber\\
&&-\sum_{p=0}^{n-1}*\left(a_1{\cal{J}}_1+a_2{\cal{J}}_2\right)\wedge{\cal{C}}_{2p+1}(\textbf{A})\rangle\nonumber
\end{eqnarray}
where $a_1$ and $a_2$ are arbitrary constants.

As a special case, in $2n+1$ dimensions a conserved current
localized on a 2$(n-1)$-brane that is a 2-form leads to the field
equations
\begin{eqnarray}
\textbf{F}^{n-1}\wedge(\textbf{F}-j)=0.
\end{eqnarray}
This implies that two special solutions for this case are
$\textbf{F}=0$ and $\textbf{F}=j$. Hence, 2-form currents may not
change the AdS curvature on the brane, or the current itself can
define the localized curvature on the brane.

\section{Special Cases}

We now consider the couplings of gravitational currents with CS
gravities in three and five dimensions and find the exact field
equations for them. These will give the local curvatures on the
branes that are induced by gravitational currents constructed from
the hidden symmetries of the global spacetime.

\subsection{Brane couplings in three dimensions}

In three dimensions the CS gravity action is equivalent to the
three-dimensional Einstein gravity with a cosmological constant. The
CS action with a coupling term is written in this case as
\begin{eqnarray}
I_3=\int_{M^3}\langle{\cal{C}}_3(\textbf{A})-j_{(2)}\wedge{\cal{C}}_1(\textbf{A})\rangle
\end{eqnarray}
where ${\cal{C}}_3(\textbf{A})=\frac{1}{2}(\textbf{A}\wedge
d\textbf{A}+\textbf{A}\wedge\textbf{A}\wedge\textbf{A})$ and
${\cal{C}}_1(\textbf{A})=\textbf{A}$. Hence the action is
\begin{eqnarray}
I_3&=&\int_{M^3}\langle\frac{1}{2}(\textbf{A}\wedge
d\textbf{A}+\textbf{A}\wedge\textbf{A}\wedge\textbf{A})-j_{(2)}\wedge\textbf{A}\rangle
\end{eqnarray}
and the corresponding field equations are
\begin{eqnarray}
\textbf{F}=j_{(2)}.
\end{eqnarray}
In source-free regions, this equation reduces to $\textbf{F}=0$ and
this implies that the spacetime has a global AdS structure,
$R^{ab}=-\frac{1}{l^2}e^a\wedge e^b$.

For the first KY current ${\cal{J}}_1$, the use of Eq. (20)
transforms the field equations (27) into
\begin{eqnarray}
\left(R^{ab}+\frac{1}{l^2}e^a\wedge
e^b\right)\textbf{J}_{ab}=-2*\left(i_{X_c}\omega_{(1)}\wedge
P^c_{AdS}\right)^{ab}\textbf{J}_{ab}.
\end{eqnarray}
In three dimensions, curvature 2-forms can be written in terms of
Ricci 1-forms and the curvature scalar \cite{Benn Tucker},
\begin{eqnarray}
R^{ab}=\frac{1}{2}{\cal{R}}e^b\wedge e^a+P^a\wedge e^b-P^b\wedge
e^a.
\end{eqnarray}
Hence the field equations are written in the form
\begin{eqnarray}
&&\left(P^a\wedge e^b-P^b\wedge e^a-\frac{1}{2}\left({\cal{R}}-\frac{2}{l^2}\right)e^a\wedge e^b\right)\textbf{J}_{ab}\nonumber\\
&=&\left[\left((i_{X_c}\omega_{(1)})i_{X_l}i_{X_k}*P^c_{AdS}\right)e^k_{AdS}\wedge
e^l_{AdS}\right]^{ab}\textbf{J}_{ab}\nonumber
\end{eqnarray}
where the equality in $n$ dimensions
$(-1)^{n-1}(*\phi)\wedge\widetilde{X}=*i_X\phi$ is used for an
arbitrary form $\phi$ and $\widetilde{X}$ is the 1-form that is the
metric dual of the vector field $X$. This can be written more
compactly as
\begin{eqnarray}
&&\left(P^{[a}\wedge e^{b]}-\frac{1}{2}\left({\cal{R}}-\frac{2}{l^2}\right)e^a\wedge
e^b\right)\textbf{J}_{ab}\nonumber\\
&=&\left[\left(\epsilon_{ckl}(i_{\widetilde{\omega_{(1)}}}P^c_{AdS})\right)e^k_{AdS}\wedge
e^l_{AdS}\right]^{ab}\textbf{J}_{ab}\nonumber
\end{eqnarray}
where $[\quad]$ on the indices denotes antisymmetrization and
$\epsilon_{ckl}$ is the completely antisymmetric Levi-Civita symbol.

Curvature 2-forms of the global AdS spacetime are
$R^{ab}_{AdS}=-\frac{1}{l^2}e^a\wedge e^b$, and the Ricci 1-forms
and the curvature scalar are obtained as
\begin{eqnarray}
P^a_{AdS}&=&-\frac{2}{l^2}e^a\nonumber\\
{\cal{R}}_{AdS}&=&-\frac{6}{l^2}\nonumber
\end{eqnarray}
and from the relation (16) the Hodge dual of the KY current
${\cal{J}}_1$ reduces to
\begin{eqnarray}
*{\cal{J}}_1=\frac{2}{l^2}*\omega_{(1)}
\end{eqnarray}
in AdS spacetime.

Let us write the KY 1-form $\omega_{(1)}$ in the co-frame basis as
follows:
\begin{eqnarray}
\omega_{(1)}=\alpha e^0_{AdS}+\beta e^1_{AdS}+\gamma e^2_{AdS}
\end{eqnarray}
where $\alpha$, $\beta$, and $\gamma$ are functions determined from
the KY equation (11) for the AdS background. By using this
definition and the curvature characteristics of AdS spacetime in Eq.
(28), the field equations in three dimensions are as follows:
\begin{eqnarray}
R^{01}+\frac{1}{l^2}e^0\wedge e^1&=&-\frac{2\gamma}{l^2}e^0_{AdS}\wedge e^1_{AdS}\nonumber,\\
R^{02}+\frac{1}{l^2}e^0\wedge e^2&=&\frac{2\beta}{l^2}e^0_{AdS}\wedge e^2_{AdS},\\
R^{12}+\frac{1}{l^2}e^1\wedge
e^2&=&\frac{2\alpha}{l^2}e^1_{AdS}\wedge e^2_{AdS}\nonumber.
\end{eqnarray}
Hence the local curvature around the brane is determined from the KY
1-forms of the global AdS spacetime. Curvature 2-forms of the brane
that differ from AdS are given as corrections to the AdS background
by KY form components.

KY forms of the three-dimensional AdS spacetime are given in
Appendix A. Let us take the KY 1-form $\omega_3$ in Eq. (A7) as an
example. Then the field equations are written as
\begin{eqnarray}
&&R^{01}+\frac{1}{l^2}e^0\wedge e^1\nonumber\\
&&=\frac{2\kappa}{l^2}\left(\frac{r^2}{l^2}+1\right)^{1/2}\cosh{(\kappa t)}\sin{\phi}dt\wedge dr,\nonumber\\
\quad\nonumber\\
&&R^{02}+\frac{1}{l^2}e^0\wedge e^2\nonumber\\
&&=\frac{2\kappa r}{l^2}\left(\frac{r^2}{l^2}+1\right)^{1/2}\sinh{(\kappa t)}\cos{\phi}dt\wedge d\phi,\\
\quad\nonumber\\
&&R^{12}+\frac{1}{l^2}e^1\wedge
e^2\nonumber\\
&&=\frac{2r^2}{l^4}\left(\frac{r^2}{l^2}+1\right)^{-1/2}\cosh{(\kappa
t)}\cos{\phi}dr\wedge d\phi\nonumber
\end{eqnarray}
and the solutions of these equations give the local co-frame on the
worldvolume of the brane. In three dimensions, only 0-branes can
couple consistently with CS theories, as can be seen from the action
(25). The wordline of the 0-brane is one dimensional and the
solutions of the field equations-namely Eq. (33)-give the geometric
structure of the wordline originating from the currents on the brane
defined from the symmetries of the global spacetime. All KY 1-forms
define conserved currents on 0-branes, and we have six different
possibilities for constructing a current. Different currents induce
different localized curvatures around the branes.

For the second current ${\cal{J}}_2$, the field equations are
changed only by a constant factor, the reason being that the main
difference coming from ${\cal{J}}_2$ is the addition of a scalar
curvature term that is constant for the AdS spacetime, as can be
seen from Eq. (17).

In fact, there is one more possible way to construct a conserved
current using two different (or identical) KY forms. From the
conservation properties of $*{\cal{J}}_1$ and $*{\cal{J}}_2$, it can
be seen that the following $(2n-(p+q))$-form in $n$ dimensions is
also a conserved current:
\begin{eqnarray}
{\cal{K}}_{(2n-(p+q))}=*{\cal{J}}_i(\omega_{(p)})\wedge*{\cal{J}}_j(\omega_{(q)}')
\end{eqnarray}
where $\omega_{(p)}$ and $\omega_{(q)}'$ are two different (or
identical) KY forms and $i,j=1,2$. In three dimensions, this current
is written as follows:
\begin{eqnarray}
{\cal{K}}_{(2)}=*{\cal{J}}_1(\omega_{(2)})\wedge*{\cal{J}}_1(\omega_{(2)}').
\end{eqnarray}
Hence, in the construction procedure of gravitational conserved
currents in three dimensions, KY 2-forms can also be used in
addition to KY 1-forms. By taking two KY 2-forms as
\begin{eqnarray}
\omega_{(2)}&=&\rho e^0_{AdS}\wedge e^1_{AdS}+\epsilon
e^0_{AdS}\wedge e^2_{AdS}+\mu e^1_{AdS}\wedge e^2_{AdS}\nonumber\\
\omega_{(2)}'&=&\nu e^0_{AdS}\wedge e^1_{AdS}+\kappa e^0_{AdS}\wedge
e^2_{AdS}+\lambda e^1_{AdS}\wedge e^2_{AdS}\nonumber
\end{eqnarray}
the field equations resulting from the current (35) are obtained as
follows:
\begin{eqnarray}
R^{01}+\frac{1}{l^2}e^0\wedge e^1&=&\frac{8}{l^4}\left(\mu\kappa-\epsilon\lambda\right)e^0_{AdS}\wedge e^1_{AdS}\nonumber\\
R^{02}+\frac{1}{l^2}e^0\wedge e^2&=&\frac{8}{l^4}\left(\rho\lambda-\mu\nu\right)e^0_{AdS}\wedge e^2_{AdS}\\
R^{12}+\frac{1}{l^2}e^1\wedge
e^2&=&\frac{8}{l^4}\left(\rho\kappa-\epsilon\nu\right)e^1_{AdS}\wedge
e^2_{AdS}\nonumber
\end{eqnarray}
where $\rho, \epsilon, \mu, \nu, \kappa$, and $\lambda$ are
functions obtained from the KY equation, as in Appendix A.

Let us take the KY 2-forms as in Eqs. (A13) and (A14),
\begin{eqnarray}
\lambda_1&=&k(\cos{\phi}e^0\wedge
e^1-\frac{1}{H_1}\sin{\phi}e^0\wedge
e^2)\nonumber\\
\lambda_2&=&k'(\sin{\phi}e^0\wedge
e^1+\frac{1}{H_1}\cos{\phi}e^0\wedge e^2)\nonumber
\end{eqnarray}
where $k$ and $k'$ are constants. Then the field equations (36)
transform into
\begin{eqnarray}
R^{01}+\frac{1}{l^2}e^0\wedge e^1&=&0\nonumber\\
R^{02}+\frac{1}{l^2}e^0\wedge e^2&=&0\\
R^{12}+\frac{1}{l^2}e^1\wedge
e^2&=&\frac{8kk'}{l^4}\frac{1}{H_1}e^1_{AdS}\wedge
e^2_{AdS}\nonumber.
\end{eqnarray}
By considering the Cartesian coordinates of a four-dimensional
hyperboloid identified with the three-dimensional AdS spacetime and
using $x^1=r\cos{\phi}$, $x^2=r\sin{\phi}$ with Eq. (A3), the last
equation reduces to
\begin{eqnarray}
R^{12}+\frac{1}{l^2}e^1\wedge e^2=\frac{8kk'}{l^4}dx^1\wedge
dx^2\nonumber.
\end{eqnarray}
Hence, the equations (37) are the same as the equations for 0-brane
worldlines with electromagnetic currents in the three-dimensional
AdS spacetime \cite{Edelstein Garbarz Miskovic Zanelli2}. In fact,
by starting with purely geometrical quantities when defining
gravitational currents, we arrive at an equation that describes the
coupling of electromagnetic currents. This may indicate a relation
between electromagnetic and gravitational currents \emph{a la}
Rainich-Misner-Wheeler theory, which states that electromagnetism
can be defined in terms of pure geometry \cite{Rainich, Misner
Wheeler}. A 0-brane solution in the electromagnetic case is obtained
by defining the 0-brane as a defect produced by an angular deficit
induced by a Killing vector field of the global spacetime
\cite{Edelstein Garbarz Miskovic Zanelli2}. The solution corresponds
to the negative mass Ba\~{n}ados-Teitelboim-Zanelli (BTZ) solution
of three-dimensional gravity with a cosmological constant
\cite{Banados Teitelboim Zanelli}. Hence, if the 0-brane is defined
as a defect in the $(x^1-x^2)$ plane produced by an angular deficit
of $2\pi\sigma=\frac{8kk'}{l^4}$, then a solution of equations (37)
is found to be the BTZ black hole metric,
\begin{eqnarray}
ds^2&=&-\left((1-\sigma)^2+\frac{r^2}{l^2}\right)dt^2+\left((1-\sigma)^2+\frac{r^2}{l^2}\right)^{-1}dr^2\nonumber\\
&&+r^2d\phi^2
\end{eqnarray}
where $M=-(1-\sigma)^2$ is the negative mass of the black hole.
However, unlike the electromagnetic case, this solution is obtained
from two of the KY 2-forms of three-dimensional AdS spacetime. This
reveals that, for the case of gravitational currents, the solutions
of the equations of motion can be obtained from the hidden
symmetries of the global spacetime. By the way, in the
electromagnetic case, the constant $2\pi\sigma$ corresponds to the
electric charge of the brane, but in the gravitational case, the
constant $\frac{8kk'}{l^4}$ comes from KY 2-forms and curvature
characteristics.

There are two new properties for the solutions of the field
equations in the gravitational-currents case. Firstly,
electromagnetic currents are written as Dirac-delta singularities
and solutions are investigated by using this property; however,
gravitational currents can couple with the global spacetime by using
KY forms, and this contains a larger class of solutions like in Eq.
(33). Secondly, solutions for the electromagnetic case can be
constructed from Killing vector identifications, but in the
gravitational case branes are constructed by using the hidden
symmetries (KY forms) of the spacetime and need not to be
constructed from Killing vector identifications.

\subsection{Brane couplings in five dimensions}

In five dimensions, there are two possible ways of coupling a
conserved current with a CS form. 4-form currents and 2-form
currents can couple consistently with CS gravity.

For the coupling of a 4-form current on the brane with CS gravity,
the action is written as
\begin{eqnarray}
I_5=\int_{M^5}\langle{\cal{C}}_5(\textbf{A})-j_{(4)}\wedge\textbf{A}\rangle
\end{eqnarray}
and the field equations become
\begin{eqnarray}
\textbf{F}\wedge\textbf{F}=j_{(4)}.
\end{eqnarray}
In the regions exterior to the brane, the equations can be solved by
$\textbf{F}=0$, which gives the AdS spacetime. However, other
solutions that satisfy $\textbf{F}\wedge\textbf{F}=0$ can also
appear.

The first KY current leads to a 4-form current $*{\cal{J}}_1$ in
terms of KY 1-forms, and the field equations take the form
\begin{eqnarray}
&&\frac{1}{4}\left(R^{ab}+\frac{1}{l^2}e^a\wedge
e^b\right)\wedge\left(R^{cd}+\frac{1}{l^2}e^c\wedge
e^d\right)[\textbf{J}_{ab},\textbf{J}_{cd}]\nonumber\\
&&=\left[*\left(-i_{X_k}\omega_{(1)}\wedge
P^k_{G}\right)\right]^{abcd}\textbf{J}_{ab}\textbf{J}_{cd}
\end{eqnarray}
and for the KY current ${\cal{J}}_2$ the field equations are
\begin{eqnarray}
&&\frac{1}{4}\left(R^{ab}+\frac{1}{l^2}e^a\wedge
e^b\right)\wedge\left(R^{cd}+\frac{1}{l^2}e^c\wedge
e^d\right)[\textbf{J}_{ab},\textbf{J}_{cd}]\nonumber\\
&&=\left[*\left(-i_{X_k}\omega_{(1)}\wedge
P^k_{G}+{\cal{R}}_G\omega_{(1)}\right)\right]^{abcd}\textbf{J}_{ab}\textbf{J}_{cd}.
\end{eqnarray}

In the case of 2-form currents coupled with CS gravity, the action
will be
\begin{eqnarray}
I_5=\int_{M^5}\langle{\cal{C}}_5(\textbf{A})-j_{(2)}\wedge{\cal{C}}_3(\textbf{A})\rangle
\end{eqnarray}
and the field equations are
\begin{eqnarray}
\textbf{F}\wedge\textbf{F}=j_{(2)}\wedge\textbf{F}.
\end{eqnarray}
Hence the first KY current ${\cal{J}}_1$ leads to the field
equations
\begin{eqnarray}
&&\frac{1}{4}\left(R^{ab}+\frac{1}{l^2}e^a\wedge
e^b\right)\wedge\left(R^{cd}+\frac{1}{l^2}e^c\wedge
e^d\right)[\textbf{J}_{ab},\textbf{J}_{cd}]\nonumber\\
&&=\left[*\left(-i_{X_k}i_{X_l}\omega_{(3)}\wedge
R^{kl}_{G}-i_{X_k}\omega_{(3)}\wedge
P^k_{G}\right)\right]^{ab}\nonumber\\
&&\wedge\frac{1}{2}\left(R^{cd}+\frac{1}{l^2}e^c\wedge
e^d\right)[\textbf{J}_{ab},\textbf{J}_{cd}]
\end{eqnarray}
and the equations obtained from the second current ${\cal{J}}_2$ are
\begin{eqnarray}
&&\frac{1}{4}\left(R^{ab}+\frac{1}{l^2}e^a\wedge
e^b\right)\wedge\left(R^{cd}+\frac{1}{l^2}e^c\wedge
e^d\right)[\textbf{J}_{ab},\textbf{J}_{cd}]\nonumber\\
&&=\left[*\left(-i_{X_k}\omega_{(3)}\wedge
P^k_{G}+{\cal{R}}_G\omega_{(3)}\right)\right]^{ab}\nonumber\\
&&\wedge\frac{1}{2}\left(R^{cd}+\frac{1}{l^2}e^c\wedge
e^d\right)[\textbf{J}_{ab},\textbf{J}_{cd}].
\end{eqnarray}
KY 2-forms and 4-forms in five dimensions can also be used instead
of KY 1-forms and 3-forms in the construction of conserved currents;
hence from Eq. (34) we obtain
\begin{eqnarray}
{\cal{K}}_{(4)}&=&*{\cal{J}}_i(\omega_{(2)})\wedge*{\cal{J}}_j(\omega_{(4)})\nonumber\\
{\cal{K}}_{(2)}&=&*{\cal{J}}_i(\omega_{(4)})\wedge*{\cal{J}}_j(\omega_{(4)}')\nonumber.
\end{eqnarray}

In five dimensions, 0-branes and 2-branes can couple consistently
with CS theories as can be seen from the actions (39) and (43),
respectively. Solutions of the field equations give the local
geometries on the worldvolumes of the branes.

\section{Conclusion}

The generalization of the minimal coupling procedure for external
sources to $p$-brane spacetimes cannot be done by extending the
coupling term to multi-index currents and connections in non-Abelian
gauge theories. However, the coupling can be considered consistently
if one uses CS forms in the coupling term. This can be relevant
because of the Abelian gauge transformation property of the CS
forms. CS theories are defined in odd dimensions, and because of the
metric independence of the action they are topological theories. By
selecting an AdS connection as the gauge connection-which includes
co-frame and spin connection-CS theories of gravity can be
constructed in odd dimensions. Hence, the coupling of
electromagnetic conserved currents on $p$-branes and CS gravities
can be consistently considered in this fashion.

For curved backgrounds, one can construct gravitational conserved
currents by using curvature characteristics and KY forms of
spacetime. These currents depend on the degree of the KY form, and
this allows for the interpretation that they are localized on
$p$-branes. Hence, the coupling of gravitational $p$-brane currents
with CS gravities can be considered in the same manner as in the
electromagnetic case. The field equations resulting from the
coupling actions gives that the one solution is an AdS spacetime for
the spacetime regions exterior to the brane. This means that the
gravitational currents are constructed from the AdS curvature and KY
forms. Therefore, the field equations give the local curvature on
$p$-branes induced by gravitational currents.

In the three-dimensional case, the field equations tell us that the
localized curvature on branes has correction terms with respect to
the AdS background written in terms of KY form components. For a
special choice of KY 2-forms, the field equations reduce to the
equations relevant for the electromagnetic coupling case, and a
special solution corresponding to the negative-mass BTZ black hole
can be found. In the five-dimensional case, there are two different
couplings and they end up with different field equations for
different branes. However, the resulting equations also give the
localized curvature on the branes.

\begin{acknowledgments}
This work was supported in part by the Scientific and Technical
Research Council of Turkey (T\"{U}B\.{I}TAK).
\end{acknowledgments}

\begin{appendix}
\section{KY Forms of AdS Spacetime in Three Dimensions}
KY forms of a class of spherically symmetric spacetimes in four
dimensions were found in Ref. \cite{Acik Ertem Onder Vercin2} by
solving the KY equation. By direct reduction, KY forms of
three-dimensional spacetimes can also be obtained from them. The
metric tensor field of AdS spacetime in three dimensions is
\begin{equation}
ds^2_{AdS}=-\left(\frac{r^2}{l^2}+1\right)dt^2+\left(\frac{r^2}{l^2}+1\right)^{-1}dr^2+r^2d\phi^2
\end{equation}
and this can be written in a locally Lorentzian form as follows:
\begin{equation}
ds^2_{AdS}=-e^0\otimes e^0+e^1\otimes e^1+e^2\otimes e^2
\end{equation}
where
\begin{equation}
e^0=H_0dt\quad,\quad e^1=H_1dr\quad,\quad e^2=rd\phi
\end{equation}
and
\begin{equation}
H_0=\left(\frac{r^2}{l^2}+1\right)^{1/2}\quad,\quad
H_1=\left(\frac{r^2}{l^2}+1\right)^{-1/2}.
\end{equation}
In three dimensions the maximal number of KY 1-forms is six, which
can be seen from Eq. (18). The corresponding KY 1-forms of AdS
spacetime are
\begin{equation}
\omega_1=H_0e^0\nonumber
\end{equation}
\begin{equation}
\omega_2=-re^2\nonumber
\end{equation}
\quad
\begin{equation}
\omega_3=(\cos{\phi})\psi_1-\frac{\kappa}{H_1}\sinh{(\kappa
t)}\sin{\phi}e^2\nonumber
\end{equation}
\begin{equation}
\omega_4=(\cos{\phi})\psi_2-\frac{\kappa}{H_1}\cosh{(\kappa
t)}\sin{\phi}e^2
\end{equation}
\begin{equation}
\omega_5=-(\sin{\phi})\psi_1-\frac{\kappa}{H_1}\sinh{(\kappa
t)}\cos{\phi}e^2\nonumber
\end{equation}
\begin{equation}
\omega_6=-(\sin{\phi})\psi_2-\frac{\kappa}{H_1}\cosh{(\kappa
t)}\cos{\phi}e^2\nonumber
\end{equation}
where $\kappa$ is an integration constant and $\psi_1$ and $\psi_2$
are defined as follows:
\begin{equation}
\psi_1=\cosh{(\kappa t)}\frac{H_0'}{H_1}e^0+\kappa\sinh{(\kappa
t)}e^1
\end{equation}
\begin{equation}
\psi_2=\sinh{(\kappa t)}\frac{H_0'}{H_1}e^0+\kappa\cosh{(\kappa
t)}e^1
\end{equation}
and $H_0'=\frac{dH_0}{dr}$.

There are four KY 2-forms, which are obtained as
\begin{equation}
\lambda_1=\cos{\phi}e^0\wedge e^1-\frac{1}{H_1}\sin{\phi}e^0\wedge
e^2
\end{equation}
\begin{equation}
\lambda_2=\sin{\phi}e^0\wedge e^1+\frac{1}{H_1}\cos{\phi}e^0\wedge
e^2
\end{equation}
\begin{equation}
\lambda_3=-\frac{w_0}{m_1}\sinh{(w_0t)}e^0\wedge
e^2+\cosh{(w_0t)}e^1\wedge e^2
\end{equation}
\begin{equation}
\lambda_4=-\frac{w_0}{m_1}\cosh{(w_0t)}e^0\wedge
e^2+\sinh{(w_0t)}e^1\wedge e^2
\end{equation}
where $m=H_0'/rH_1$, $m_1=(r/H_0)'H_0^2/H_1$, and $mm_1=\pm w_0^2$.
In all dimensions, the volume form multiplied with a constant
automatically satisfies the KY equation. Hence, the KY 3-form in
three dimensions is
\begin{equation}
\omega_{(3)}=ce^0\wedge e^1\wedge e^2
\end{equation}
where $c$ is a constant.

\end{appendix}



\begin{references}

\bibitem{Teitelboim} C. Teitelboim, Phys. Lett. B \textbf{167}, 63
(1986).

\bibitem{Henneaux Teitelboim} M. Henneaux and C. Teitelboim, Found.
Phys. \textbf{16}, 593 (1986).

\bibitem{Deser Jackiw Templeton} S. Deser, R. Jackiw, and S.
Templeton, Ann. Phys. \textbf{140}, 372 (1982).

\bibitem{Achucarro Townsend} A. Achucarro and P. K. Townsend, Phys.
Lett. B \textbf{180}, 89 (1986).

\bibitem{Zanelli1} J. Zanelli, Classical Quantum Gravity
\textbf{29}, 133001 (2012).

\bibitem{Miskovic Zanelli} O. Miskovic and J.
Zanelli, Phys. Rev. D \textbf{80}, 044003 (2009).

\bibitem{Edelstein Garbarz Miskovic Zanelli} J. D. Edelstein, A.
Garbarz, O. Miskovic, and J. Zanelli, Int. J. Mod. Phys. D
\textbf{20}, 839 (2011).

\bibitem{Bunster Gomberoff Henneaux} C. Bunster, A. Gomberoff, and M.
Henneaux, Phys.Rev. D \textbf{84}, 125012 (2011).

\bibitem{Arnowitt Deser Misner} R. Arnowitt, S. Deser, and C. W.
Misner, in \emph{Gravitation: An Introduction to Current Research},
edited by L. Witten, (Wiley, New York, 1962).

\bibitem{Kastor Traschen} D. Kastor and J. Traschen, J. High Energy
Phys. 08 (2004) 045.

\bibitem{Acik Ertem Onder Vercin1} O. Acik, U.
Ertem, M. Onder, and A. Vercin, Gen. Relativ. Gravit. \textbf{42},
2543 (2010).

\bibitem{Acik Ertem Onder Vercin2} O. Acik, U.
Ertem, M. Onder, and A. Vercin, J. Math. Phys. \textbf{51}, 022502
(2010).

\bibitem{Nakahara} M. Nakahara, \emph{Geometry, Topology and
Physics} 2nd edition, (Taylor and Francis, London, 2003).

\bibitem{Wise} D. K. Wise, Classical Quantum Gravity \textbf{27}, 155010
(2010).

\bibitem{Barbour Pfister} J. Barbour and H. Pfister (editors), \emph{Mach's Principle: From Newton's Bucket to Quantum Gravity},
(Birkhauser, Boston, 1995).

\bibitem{Benn Tucker} I. M. Benn and R. W. Tucker, \emph{An Introduction to Spinors and Geometry with Applications in
Physics} (IOP Publishing Ltd, Bristol, 1987).

\bibitem{Edelstein Garbarz Miskovic Zanelli2} J. D. Edelstein, A.
Garbarz, O. Miskovic, and J. Zanelli, Phys. Rev. D \textbf{82},
044053 (2010).

\bibitem{Rainich} G. Y. Rainich, Trans. Am. Math. Soc. \textbf{27},
106 (1925).

\bibitem{Misner Wheeler} C. W. Misner and J. A. Wheeler, Ann. Phys.
\textbf{2}, 525 (1957).

\bibitem{Banados Teitelboim Zanelli} M. Banados, C. Teitelboim, and
J. Zanelli, Phys. Rev. Lett. \textbf{69}, 1849 (1992).

 \end{references}
\end{document}